\documentclass[
showkeys,    
reprint,    
amsmath,    
amssymb,    
amsfonts,    
aps,        
prl%
groupedaddress,
superscriptaddress,
a4paper,%
nofootinbib,
eprint,        
final,        
floatfix,    %
preprintnumbers, 
]{revtex4-2}

\usepackage{amsthm,mathtools}

\usepackage[utf8]{inputenc}
\usepackage[T1]{fontenc}

\usepackage{physics}

\usepackage{graphics,graphicx} 

\usepackage{dutchcal}

\usepackage{color,xcolor}
\definecolor{ForestGreen}{RGB}{34,139,34}      
\definecolor{RubinRed}{RGB}{224,17,95}    

\usepackage[
]{hyperref}

\usepackage{natbib}
\hypersetup{
colorlinks=true,
linkcolor=RubinRed,
filecolor=ForestGreen,      
urlcolor=ForestGreen,
citecolor=cyan,
anchorcolor=orange,
linktocpage=true,
breaklinks,
breaklinks=true,
final,
pdftitle={Morphology-resolved scrambling in a chaotic quantum billiard},
pdfauthor={Pranaya Pratik Das},
}

\usepackage{accents}

\usepackage[capitalise]{cleveref}
\crefname{figure}{FIG.}{FIGS.}
\Crefname{figure}{FIG.}{FIGS.} 
\makeatletter
\AtBeginDocument
{
	\def\ltx@label#1{\cref@label{#1}}
	\def\label@in@display@noarg#1{\cref@old@label@in@display{#1}}
	\def\label@in@mmeasure@noarg#1{%
		\begingroup%
		\measuring@false%
		\cref@old@label@in@display{#1}
		\endgroup}%
} %
\makeatother
	
	\usepackage{orcidlink}

\begin{document}
\normalsize
\title{Morphology-resolved scrambling in a chaotic quantum billiard}
\author{Pranaya Pratik \surname{Das}\,\orcidlink{0000-0002-6025-7719}}
\email{pranaya.phy@outlook.com}
\homepage{ppdws.github.io/pranayapratikdas/}
\affiliation{Department of Physics and Astronomy, National Institute of Technology Rourkela, Odisha, India-769008}

\date{\today}

\begin{abstract}
Chaotic quantum systems can retain spatial memory through scarred eigenstates, but whether these static structures control scrambling remains unclear. This work establishes a morphology-resolved connection between scarred eigenstates and eigenstate-resolved OTOCs in a peanut-shaped quantum billiard. Scalar localisation diagnostics, including differential entropy and continuum participation ratios, detect anomalous concentration but discard spatial architecture. A scale-normalised density overlap, in contrast, directly compares probability density profiles, revealing families of orthogonal eigenstates with nearly identical spatial morphology. Comparing the complete OTOC time traces of these orthogonal eigenstates reveals that morphological recurrence has dynamical content: moderate density overlap yields no universal prediction, whereas strongly recurring morphologies exhibit nearly identical OTOC growth and saturation. Thus, scarred structures act as spatial templates for operator growth, not merely static violations of ergodicity. This morphology-resolved framework turns eigenstate shape into a quantitative predictor of scrambling and provides a scale-controlled diagnostic of weak ergodicity breaking in quantum chaos.
	\end{abstract}
	
	
	\maketitle
	\pagestyle{plain}
	
	\paragraph*{\label{sec:1}Introduction---}
	Chaotic quantum systems do not always erase memory of their initial conditions; instead, a sparse set of nonthermal eigenstates can emerge throughout the spectrum. A central challenge is therefore not only to identify the origin of these nonthermal structures, but also to classify the recurring morphologies and determine whether these static structures leave measurable signatures in the subsequent dynamics.

	The \textit{Eigenstate Thermalisation Hypothesis} (ETH) suggests that individual eigenstates of nonintegrable many-body systems should behave thermally\cite{rigol2008thermalization,d2016quantum}. This implies that local observables will equilibrate and lose memory of their initial conditions. In single-particle chaotic billiards, the analogous expectation appears through quantum ergodicity and random-wave statistics\cite{PhysRevLett.53.1515,Turner2018}. However, significant deviations from this prediction are evident. Quantum scars are a prime example of such deviations. These are anomalous nonthermal eigenstates whose probability densities remain enhanced near slowly diverging classical trajectories\cite{das2026classicalquantumchaosbean, PhysRevLett.134.140402, graf2024birthmarksergodicitybreakingquantum}, leading to weak ergodicity breaking across a predominantly chaotic spectrum. 

	Scarring, however, is not just a question of localisation strength. Quantifying their existence, recurrence across the spectrum, and determining their dynamical consequences remains nontrivial. Standard scar diagnostics, including overlaps with selected classical trajectories, Husimi projections, participation ratios, entropy measures, and semiclassical constructions, have been highly successful. However, they often rely on prior knowledge of the relevant orbit, phase-space representation, or system geometry. On top of these dependencies, they do not determine whether different eigenstates share the same spatial morphology, nor whether such morphological similarity leads to similar scrambling dynamics. The Shannon differential entropy\cite{michalowicz2013handbook, 1055832} and the \textit{Inverse Participation Ratio}\cite{PhysRevB.107.235108} (IPR), both members of the Rényi entropy hierarchy, are widely used to quantify how anomalously localised a continuous PDF $\rho(\mathbf{x})(=\abs{\psi(\mathbf{x})}^{2})$ is\cite{michalowicz2013handbook,1055832,PhysRevB.107.235108}. These quantities must nevertheless be interpreted with care, since differential entropy depends on the choice of coordinates and length units, whereas the continuum IPR is dimensional and sensitive to normalisation, domain size, and spatial resolution. These measures are therefore useful indicators of concentration alone.

	To overcome these limitations, we use scale-normalised and relative diagnostics. In particular, we compare the probability density of each eigenstate with a carefully constructed uniform probability density distribution through relative measures such as the \textit{Kullback-Leibler} (KL) divergence and the \textit{normalised Partition Ratio} (nPR). Then we introduce a pair-wise normalised density overlap that directly compares the full probability density profiles of distinct eigenstates. This allows eigenstates to be grouped into morphology classes ($\mathcal{G}_{\mathtt{g}}$) without assuming in advance which classical orbit, if any, is responsible for the structure.

	To test whether these static classes have dynamical significance, we compute eigenstate-resolved \textit{Out-of-Time-Order Correlators} (OTOCs)\cite{Larkin, Hashimoto2020, GarcaMata2023, Das2025, PhysRevLett.126.030601, das2026classicalquantumchaosbean}, which quantify the growth of operator noncommutativity and the scrambling of initially local information. For each pair of eigenstates, we compare the complete OTOC time traces through a normalised curve distance and its associated similarity. This directly asks whether near-identical density morphologies lead to near-identical scrambling responses.

\begin{figure}[tbh!]
		\centering
		\includegraphics[width=\linewidth]{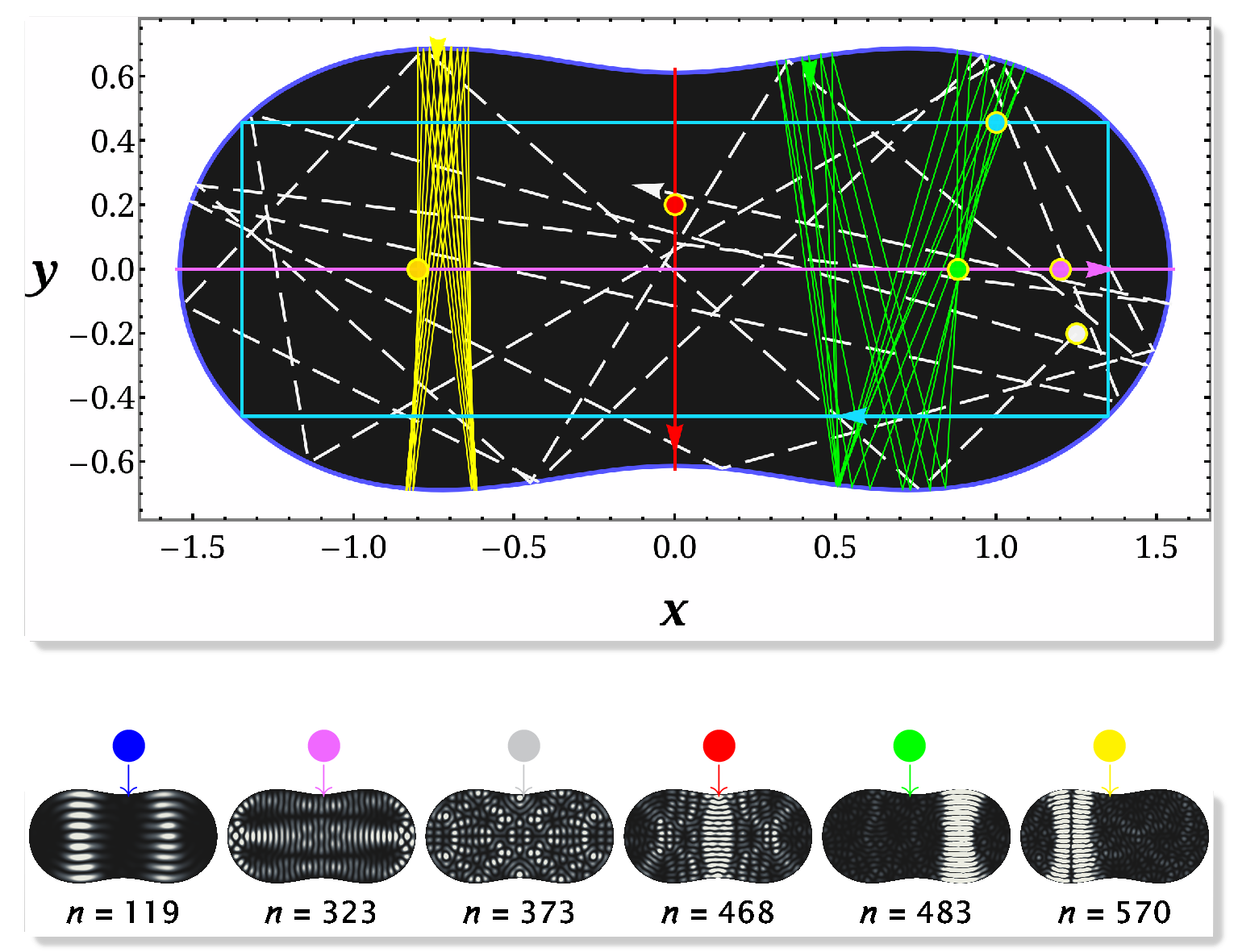}
		\caption{Classical trajectories and eigenstate intensities in the peanut-shaped billiard. \textbf{Top:} coloured curves mark selected slowly diverging orbits, while the white dashed curve indicates chaotic motion. \textbf{Bottom:} eigenstate intensities $\abs{\psi(\mathbf{x})^{2}}$ show enhanced probability density along the corresponding classical structures. The colour coding connects each eigenstate morphology to the corresponding classical structure in the upper panel.}
		\label{Fig:1}
	\end{figure}

	We apply this framework to the peanut-shaped quantum billiard\cite{das2026classicalquantumchaosbean}, whose boundary ($\Omega_{\mathrm{peanut}}$) is defined by $(x^{2}+y^{2})^{2}-2 (x^{2}-y^{2})+1-1.375^{2}=0$. Its mixed-curvature geometry supports unstable periodic orbits embedded in a predominantly chaotic phase space, making it a minimal two-dimensional fertile ground for exploring scar formation. Representative scarred eigenstates and their associated slowly diverging classical trajectories are shown in \cref{Fig:1}. In this article, we show that morphology-resolved density overlap identifies recurring scar families across the spectrum and that these families exhibit closely related OTOC dynamics. Thus, scar morphology is not merely a static feature of anomalous eigenfunction; it provides a predictive link between nonthermal eigenstate structure and quantum information scrambling.

	\paragraph*{\label{sec:2}Relative entropy---}
	For a normalised eigenstate $\psi_{n}(\mathbf{x})$ on the billiard domain $\omega_{\textrm{peanut}}$, the spatial spread of $\rho_{n}(\mathbf{x})$ is provided by the Rényi entropy, a one-parameter family of measures, defined as
	\begin{equation}\label{Eq:1}
		H_{\alpha}\Big(\rho_{n}(\mathbf{x})\Big):=\frac{\log\Big(\int_{\Omega}\rho_{n}(\mathbf{x})^{\alpha}\dd{\mathbf{x}}\Big)}{1-\alpha}, \quad \alpha\in (0,1)\cup(1,\infty)
	\end{equation}
	The parameter $\alpha$ controls which part of the density is emphasised: $\alpha>1$ gives stronger weight to high-density regions, while $\alpha<1$ is more sensitive to low-density support. In the limit $\alpha\to1$, \cref{Eq:1} gives the differential Shannon entropy:
	
	\begin{equation}\label{Eq:2}
	H_{S}=-\int_{\Omega}\rho_{n}(\mathbf{x})\log\Big(\rho_{n}(\mathbf{x})\Big)\dd{\mathbf{x}}
	\end{equation}
	Large $H_{S}$ indicates a spatially extended state, whereas a reduced value signals a localised state.
	
	A related localisation measure is the IPR\cite{PhysRevB.107.235108},
	\begin{equation}\label{Eq:3}
	\textrm{IPR}=\int_{\Omega}\rho_{n}(\mathbf{x})^{2}\dd{\mathbf{x}}
	\end{equation}
	It is the second moment of the probability density and corresponds to the Rényi entropy at $\alpha=2$ in \cref{Eq:1} through $H_{2}=-\log(\textrm{IPR})$. Thus, sharply localised eigenstates have large IPR, whereas extended states have small IPR that scales inversely with the effective volume (or area) of the system. Intuitively, the inverse of the IPR can serve as a natural measure of delocalisation. 
	
	Both $H_{S}$ and IPR are logical extensions of their discrete counterparts, given that the probability distributions are continuous. Unlike Wehrl entropy\cite{ZYCZKOWSKI2001583, C_T_Lee_1988}, which relies on the Husimi Q function to ensure non-negativity, differential entropy directly employs the probability density function, with no built-in smoothing, making it sensitive to the units and coordinate system used and hence is not invariant under transformations\cite{e6050388}. In particular, $H_{S}$ can assume both positive or negative values\cite{statproofbookDifferentialEntropy, cover2012elements, shannon1948mathematical, e23060778} because continuous PDF ($\rho(\mathbf{x})$) is not restricted to the values less than unity\cite{griffiths2018introduction, casella2024statistical}. In a similar vein, for continuous probability distributions, if $\psi(\mathbf{x})$ is sharply peaked, then $\abs{\psi(\mathbf{x})}^{4}$ becomes large, and the resulting IPR value may surpass unity\cite{Clark_2018}. These limitations motivate the use of normalised or relative measures that remove dependence on units and allow meaningful comparisons across different states, systems, and geometries.
	
	One such relative measure is the \textit{Kullback-Leibler} (KL) divergence\cite{Joyce2025} between $\rho_{n}(\mathbf{x})$ and a reference density $\rho_{0}(\mathbf{x})$,
	\begin{equation}\label{Eq:4}
	\begin{split}
		D_{KL}(\rho_{n}||\rho_{0})&:= \int_{\Omega} \rho_{n}(\mathbf{x})\log(\frac{\rho_{n}(\mathbf{x})}{\rho_{0}(\mathbf{x})})\dd{\mathbf{x}}\\
		&= \int_{\Omega} \rho_{n}(\mathbf{x})\log{\big(\mathcal{A} \times \rho_{n}(\mathbf{x})\big)}\dd{\mathbf{x}}
	\end{split}
	\end{equation}
	where $\rho_{0}$ ($=\dfrac{1}{\mathcal{A}}$) is the referenced uniform density distribution over a bounded two-dimensional domain $\Omega$ of area $\mathcal{A}$ $\big(=\int_{\Omega}\dd{\mathbf{x}}\big)$. Thus, $D_{KL}(\rho_{n}||\rho_{0})$ measures the deviation of an eigenstate from the uniform distribution. It vanishes for a perfectly uniform density and increases as the probability weight becomes more spatially concentrated in a sub-region.
	
	\begin{figure*}[tbh!]
	\centering
	\includegraphics[width=\linewidth]{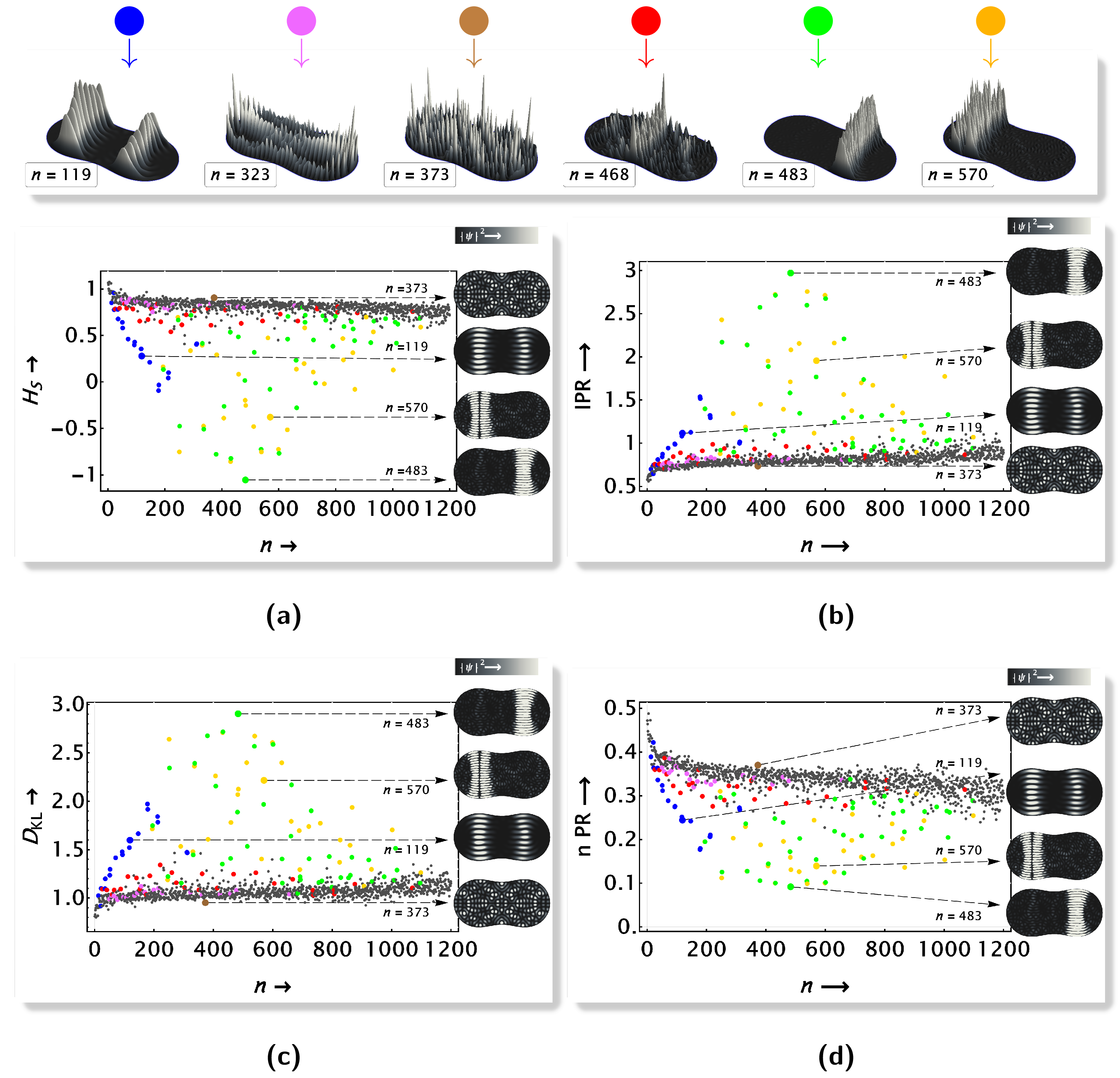}
	\caption{Eigenstate morphology and scar diagnostics across the spectrum. \textbf{Top:} representative intensity profiles $\abs{\psi}^{2}$, with colours matching the highlighted data points. \textbf{Bottom:} spectral variation of (a) Differential Shannon entropy $H_{S}$, (b) IPR, (c) $D_{\textrm{KL}}$, and (d) nPR versus eigenstate index $n$. Insets display a few selected eigenstate probability densities. While these scalar diagnostics reveal anomalous concentrations and deviations from ergodic spatial filling, they collapse each eigenstate to a single number and therefore cannot resolve whether distinct eigenstates share the same spatial morphology.}
	\label{Fig:2}
\end{figure*}

	We also use the \textit{normalised Participation Ratio} (nPR), 
	\begin{equation}
		\textrm{nPR}=\frac{\rho_{0}}{\textrm{IPR}}=\frac{1}{\mathcal{A} \times \textrm{IPR}}\quad \in [0, 1]
	\end{equation}
	Here, nPR is the exponential of Rényi entropy of order $2$, normalised by the domain size $\mathcal{A}$. The value $\mathrm{nPR}=1$ corresponds to a uniform density over the entire billiard domain, while smaller values indicate that the eigenstate effectively occupies only a fraction of the available area. Unlike the raw IPR, nPR is dimensionless and directly comparable across different system sizes.
	
	\cref{Fig:2} summarises these diagnostics across the spectrum. The raw quantities $H_{S}$ and IPR, in \cref{Fig:2}(a \& b), identify states with enhanced spatial concentration, including scarred or partially localised eigenstates. However, their raw absolute values are scale-dependent. For example, for a strongly concentrated eigenstate such as for $n=483$, $H_{S}$ acquires a negative value, and IPR exceeds unity for the same state. These features do not invalidate the localisation trend, but they make absolute comparisons across states less transparent and create interpretational difficulties.
	
	\cref{Fig:2}(c \& d) present dimensionless diagnostics that remove this ambiguity by using dimensionless, scale-independent measures. The KL divergence highlights deviations from the uniform density, so localised or scarred eigenstates appear as large positive outliers. The nPR rescales the participation ratio by the billiard area and therefore measures the fraction of the domain effectively occupied by the eigenstate. Scarred or localised states have comparatively large $H_{S}$ and small nPR, while extended states show the opposite trend. Taken together, these measures separate genuine spatial concentration from unit-dependent artefacts.
	
	\paragraph*{Pairwise density overlap and morphology classes---}
	\begin{figure*}[tbhp!]
		\centering
		\includegraphics[width=\linewidth]{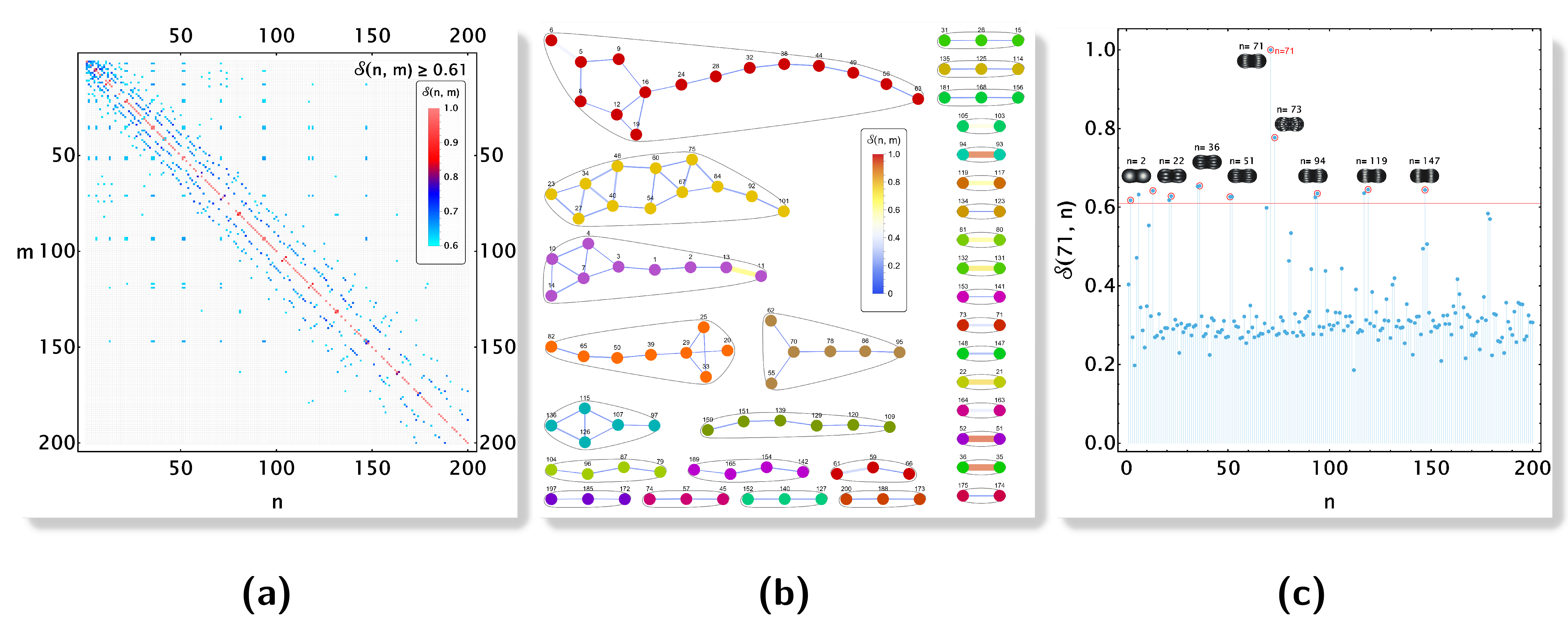}
		\caption{Density-overlap similarity between eigenstates. (a) Thresholded matrix $\mathcal{S}(n,m)\geq0.61$, where off-diagonal elements near unity identify distinct eigenstates with nearly identical probability densities. (b) Network of spatially similar eigenstates, with edges drawn for $\mathcal{S}(n,m)\geq0.67$; edge colour and thickness encode $\mathcal S(n,m)$. The balloon layout is only visual. (c) Eigenstates most similar to the reference state $n=71$; the red line marks $\mathcal{S}_{\rm th}=0.61$.}
		\label{Fig:3}
	\end{figure*}
	
	Localisation strength alone, however, does not determine whether two scarred states have the same spatial structure. To compare morphology directly, we introduce the $L_{2}$-normalised\footnote{An $L_{2}$-normalised density profile is defines as $\tilde{\rho}(\mathbf{x})=\frac{\rho(\mathbf{x})}{\sqrt{\int_{\Omega}\rho^{2}(\mathbf{x})\dd{\mathbf{x}}}}$, such that $\norm*{\tilde{\rho}(x)}_{2}^{2}=1$. Unlike standard $L_{1}$ normalisation, $L_{2}$ normalisation makes the integral of the squared height equal to $1$. This suppresses peaks and emphasises tails, allowing direct comparison of different spatial distributions based on their root‑mean‑square scales.} density overlap
	\begin{equation}\label{Eq:6}
		\mathcal{S}(n,m)=
		\frac{
			\int_{\Omega}\bigg(\rho_{n}(\mathbf{x}) \times \rho_{m}(\mathbf{x})\bigg)\dd{\mathbf{x}}}
		{\sqrt{\bigg(
				\int_{\Omega}\rho_{n}^{2}(\mathbf{x})\dd{\mathbf{x}}\bigg) \times
				\bigg(\int_{\Omega}\rho_{m}^{2}(\mathbf{x})\dd{\mathbf{x}}
				\bigg)}}	
	\end{equation} 
	By construction, $0\le \mathcal{S}(n,m)\le 1$, with $\mathcal{S}(n,n)=1$ (diagonal elements in the matrix plot reflect the trivial self-similarity as shown in \cref{Fig:3}(a)). Thus, $\mathcal{S}(n,m)\simeq 1$ implies
	\begin{equation}\label{Eq:7}
		\norm*{\tilde{\rho}_{n}-\tilde{\rho}_{m}}_{2}^{2}=2(1-\mathcal{S}(n,m))\ll 1.
	\end{equation}
	 That means the two eigenstates have almost identical spatial intensity fingerprints. They place their probability weight in nearly the same sub-regions of the billiard, whereas $\mathcal{S}(n,m)\ll 1$ indicates that the two states occupy different sub-regions or have strongly different intensity patterns. 
	
	The numerator in \cref{Eq:6} is a cross-participation or an intensity-overlap integral. For $n=m$, it reduces to IPR in \cref{Eq:3}. Therefore, $\mathcal{S}(n,m)$ can be interpreted as a normalised cross-participation ratio: it measures the degree to which two eigenstates concentrate their probability weight in the same spatial regions, after removing the trivial effect of different individual localisation strengths. This is distinct from the Hilbert-space overlap $\braket{n}{m}$, which vanishes for different eigenstates of a Hermitian Hamiltonian. Two orthogonal eigenfunctions can therefore have $\braket{n}{m}=0$, but still display a large density overlap for $\mathcal{S}(n,m)\approx1$.
	
	A coarse-grained interpretation makes this connection transparent. Let
	\begin{equation}
		\begin{split}
		\hat P_{\mathcal{a}}&=\int_{\Omega_{\mathcal{a}}}\ket{\mathbf{x}}\bra{\mathbf{x}}\dd{\mathbf{x}},\\
		p_{n\mathcal{a}}&=\mel{n}{\hat P_{\mathcal{a}}}{n}
			=\int_{\Omega_{\mathcal{a}}}\rho_{n}(\mathbf{x})\dd{\mathbf{x}} ,
		\end{split}
	\end{equation}
	where $\Omega_{\mathcal{a}}$ is a coarse-graining of the billiard. The vector $\textbf{p}_{n}=(p_{n1},p_{n2},\ldots)$ gives a discrete representation of the local spatial weight of the eigenstate $n$ such that $\sum_{\mathcal{a}}p_{n\mathcal{a}}=1$. In this representation, $\mathcal{S}(n,m)=\dfrac{\mathbf{p}_n\cdot \mathbf{p}_m} {\norm*{\mathbf{p}_n}\norm*{\mathbf{p}_m}}$. In the continuum limit, this reduces to \cref{Eq:6}. So, $\mathcal{S}(n,m)$ is nothing but the cosine similarity\footnote{Cosine similarity is the normalised inner product of two vectors.} between two density vectors. Hence, $\mathcal{S}(n,m)$ does not compare the amplitudes of $\psi_{n}$ and $\psi_{m}$ directly; it compares their observable spatial probability distributions.

	Hence, $\mathcal{S}(n, m)$ is a configuration-space diagnostic. It is insensitive to the overall phase or sign of an eigenfunction, which is useful for comparing standing-wave patterns. It does not resolve momentum direction or phase-space structure.
	
	\cref{Fig:3} shows that eigenstate similarity is not distributed randomly across the spectrum but organises into structured families. The thresholded density-overlap matrix, $\mathcal{S}(n,m)\geq 0.61$, in \cref{Fig:3}(a) shows that similarity is not restricted to the diagonal ($n=m$), but appears in structured off-diagonal bands and isolated high-overlap pairs, indicating distinct eigenstates with comparable spatial intensity profiles. In \cref{Fig:3}(b), recasts these high-overlap pairs as a network, where vertices denote eigenstates and edges connect states with $\mathcal{S}_{th}=0.67$, a stricter threshold chosen for graph clarity. The connected components identify families of eigenstates with similar density morphology, while edge colour and width encode the magnitude of $\mathcal{S}(n,m)$. Finally, (\cref{Fig:3}(c)) shows the similarity profile relative to $n_{0}=71$, identifying states whose densities closely resemble the reference state. Together, \cref{Fig:3} reveals recurring spatial structures embedded in the eigenstate sequence. This morphological information is dynamically relevant because local operators probe the regions where an eigenstate carries probability weight.
	
	\paragraph*{\label{sec:6}Scrambling dynamics and scarred multiplets---}
	Comparing eigenstate densities with eigenstate-resolved (microcanonical) OTOCs provides a direct test of whether static morphology has dynamical consequences\cite{maldacena2016bound}. For local Hermitian operators $\hat{W}$ and $\hat{V}$ the squared commutator
		\begin{equation}\label{Eq:9}
		C_{n}(t):=\mel{n}{\comm{\hat{W}(t)}{\hat{V}}^{\dagger}\comm{\hat{W}(t)}{\hat{V}}}{n}
	\end{equation} 
	can be written as a combination of time-ordered and out-of-time-ordered 4-point correlators
	, which depend on products of local matrix elements. In the Heisenberg picture, 
	\begin{equation}\label{Eq:11}
		\mel{a}{\hat{W}(t)}{b}=\emph{e}^{\frac{\imath t}{\hbar}(\mathcal{E}_{a}-\mathcal{E}_{b})} \hat{W}_{ab}, \quad  \because \hat{W}_{ab}=\mel{a}{\hat{W}}{b},
	\end{equation}
	the 4-point OTOC has the energy-basis representation
	\begin{equation}\label{Eq:12}
		C_{n}(t)=\sum_{a,b,c}\emph{e}^{\frac{\imath t}{\hbar}(\mathcal{E}_{n}-\mathcal{E}_{a}+\mathcal{E}_{b}-\mathcal{E}_{c})}\hat{W}_{na}\hat{V}_{ab}\hat{W}_{bc}\hat{V}_{cn}
	\end{equation}
	
	This expression shows that eigenstate-resolved OTOCs are governed by three ingredients: energy gaps, which generate oscillatory phases; operator matrix elements, which determine the couplings induced by $\hat{W}$ and $\hat {V}$; and eigenfunction structure, which controls the spatial organisation of those matrix elements.

	For the Hermitian pair $\{\hat{x}(t),\hat{p}\}$, the rapid growth of $C_{n}(t)$ measures the increasing noncommutativity. In a semiclassical chaotic system, its early-time growth reflects local phase-space stretching and can scale as $C_{n}(t)\sim e^{2\lambda_{L} t}$ before finite-size saturation, where $\lambda_L$ is the quantum Lyapunov exponent. Thus, eigenstates in the same high-$\mathcal{S}(n,m)$ morphology class are sampled by $\hat{x}(t)$ and $\hat{p}$ over nearly the same regions of the billiard. Their local matrix-element products are therefore correlated, producing similar short-to-intermediate-time OTOC dynamics. In this sense, recurring spatial morphology constrains how local information is scrambled.

	\begin{figure}[tbh!]
	\centering
	\includegraphics[width=\linewidth]{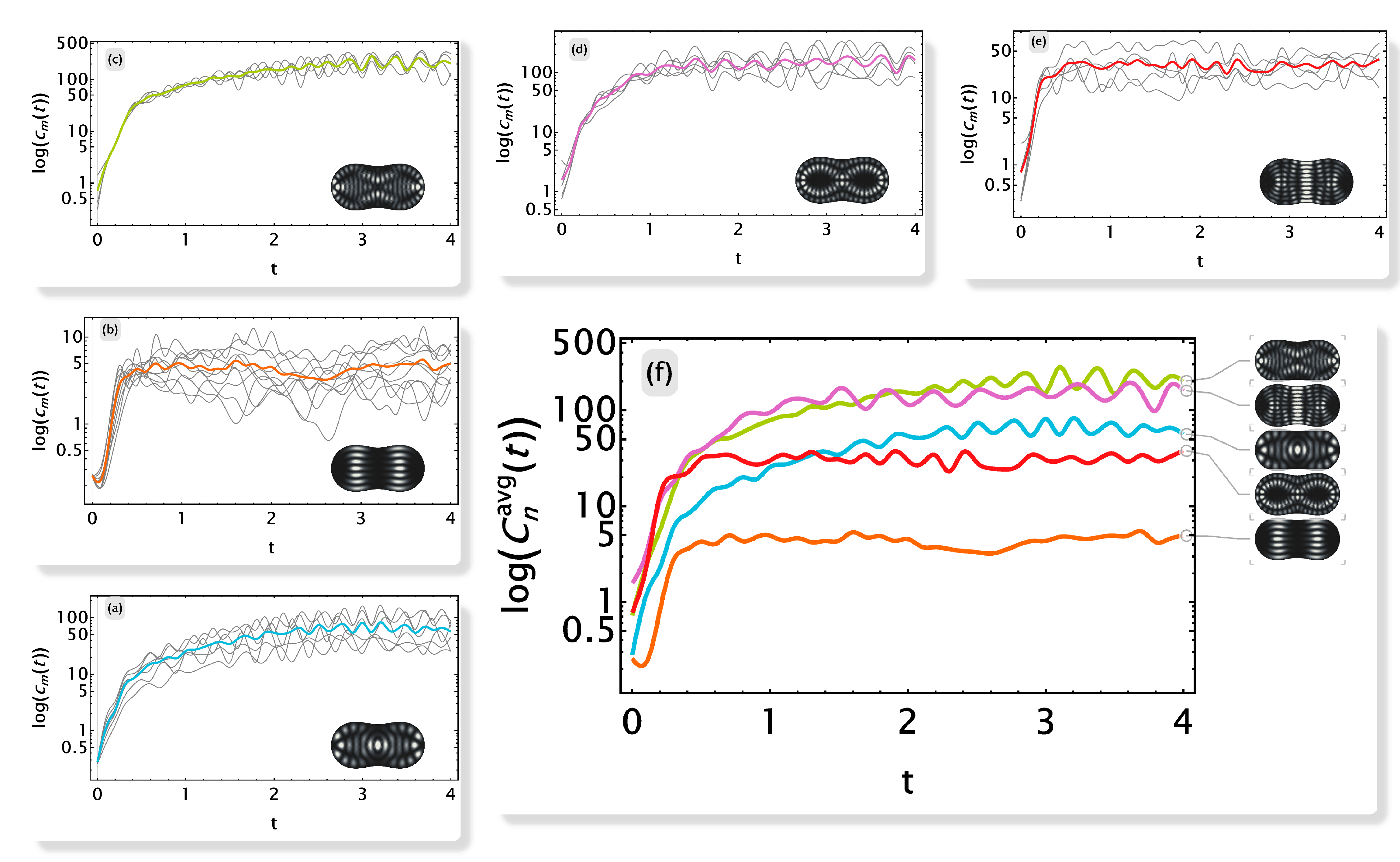}
	\caption{Morphology-resolved OTOC dynamics. (a-e) show eigenstate groups $\mathcal{G}_{\mathtt{g}}$ with similar density morphology, identified by the overlap $\mathcal{S}(n,m)$. Grey curves are individual OTOCs and coloured curves are group averages $C_{\mathtt{g}}^{\mathrm{avg}}(t)$. Density thumbnails show representative members from the corresponding morphology class. (f) compares averaged curves across all classes, demonstrating that recurring eigenstate morphologies yield distinct yet internally consistent OTOC growth and saturation profiles.}
	\label{Fig:4}
	\end{figure}

	This provides the physical basis for morphology-resolved scrambling. We first identify morphology classes $\mathcal{G}_{\mathtt{g}}$ as groups of eigenstates connected by large $\mathcal{S}(n,m)$. Its representative scrambling profile is quantified by the class-averaged OTOC $C_{\mathtt{g}}^{\mathrm{avg}}(t)$   $(=\frac{1}{\abs{\mathcal{G}_{\mathtt{g}}}}\sum_{n\in \mathcal{G}_{\mathtt{g}}} C_{n}(t))$. As shown in \cref{Fig:4}, classes with similar intensity morphology display comparable OTOC growth and saturation behaviour. Thus recurring spatial structures are not merely static features of the eigenfunctions, but also organise the dynamical response.

	To quantify this connection pair-wise, we compare the complete OTOC time traces through the normalised OTOC distance between two eigenstates $n$ and $m$:
	\begin{equation}\label{Eq:12}
		d_{c}(n,m)=
		\frac{
			\sqrt{\int_{0}^{t}\abs{C_{n}(t)-C_{m}(t)}^{2}\dd{t}}}
		{\sqrt{\int_{0}^{t}\abs{C_{n}(t)}^{2}\dd{t}} +
			\sqrt{\int_{0}^{t}\abs{C_{m}(t)}^{2}\dd{t}}}
	\end{equation}
	For uniformly sampled data, the integrals are replaced by sums with time step $\Delta t$. This distance compares the full-time traces directly, without fitting an exponential growth rate or extracting a single characteristic time. Therefore, it is sensitive to differences in early growth, oscillations, saturation, and revivals.
	
	By the triangle inequality, $0 \leq d_{c}(n, m) \leq 1,$ provided the denominator is nonzero. Thus, $d_{c}(n, m) \simeq 0$ implies $C_{n}(t) \simeq C_{m}(t)$, i.e. nearly identical OTOC profiles, while $d_{c}\simeq1$ indicates strongly different profiles. It is then useful to define an OTOC-curve similarity: $ Q_{c}(n,m)=1-d_{c}(n,m)$, so that $Q_{c}(n,m)\approx1$ corresponds to nearly identical scrambling dynamics. In contrast, a small value corresponds to dissimilar dynamics. 
	
	\begin{figure}[hbt!]
		\centering
		\includegraphics[width=\linewidth]{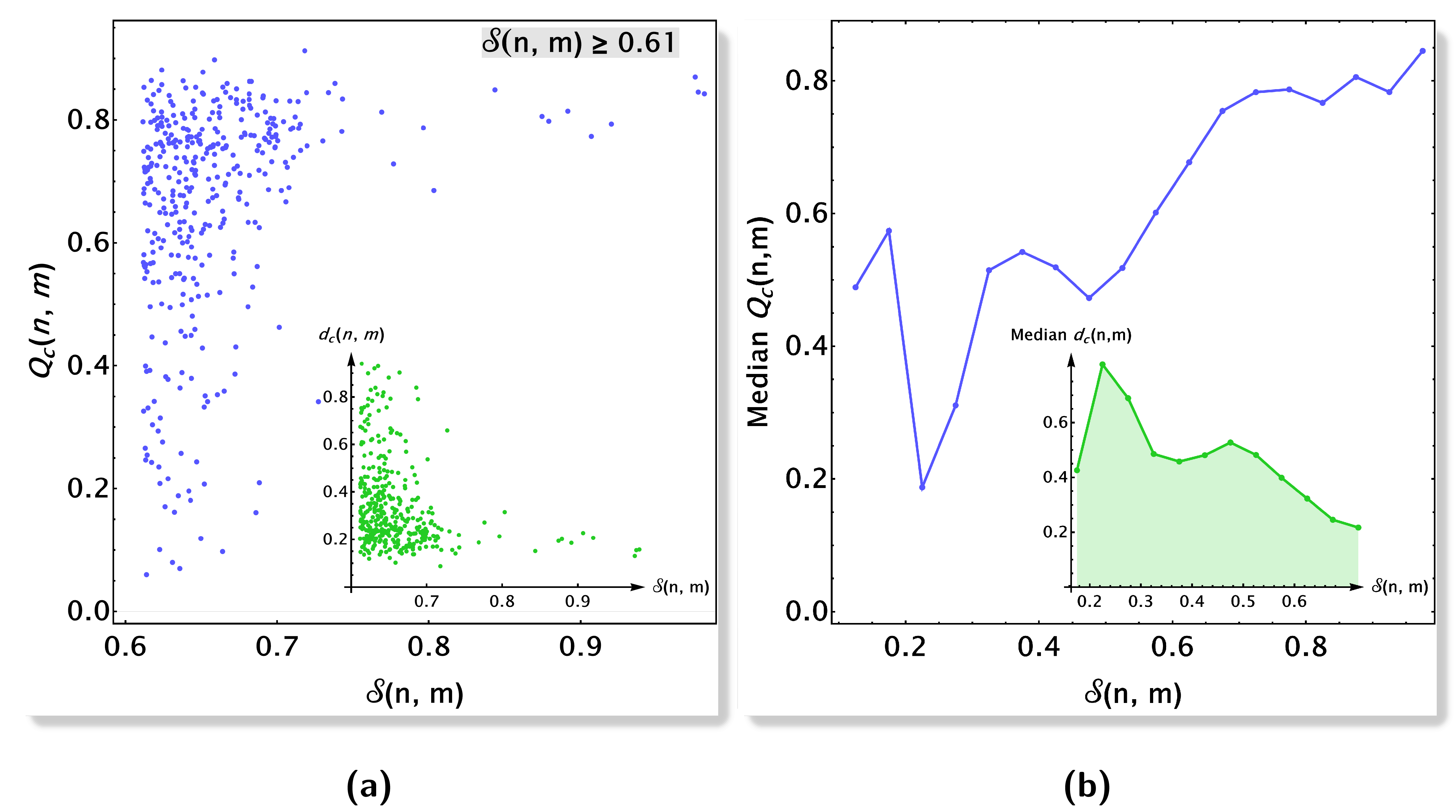}
		\caption{Pair-wise relation between spatial morphology and OTOC. (a) OTOC-curve similarity $Q_{c}(n,m)$ as a function of $\mathcal{S}(n,m)$ for high-overlap pairs satisfying $\mathcal{S}(n,m)\ge0.61$. The inset shows the complementary $ d_{c}(n,m)$ vs $\mathcal{S}(n,m)$. Pairs near the cut-off show broad dynamical variation, whereas the largest-$\mathcal{S}$ pairs are biased toward high $Q_{c}$ and low $d_{c}$. (b) Binned median OTOC similarity as a function of $\mathcal{S}(n,m)$. The systematic increase of median $Q_{c}$, together with the decrease of median $d_{c}$ shown in the inset, demonstrates that strongly overlapping spatial-density profiles lead to nearly identical OTOC time traces.}
		\label{Fig:5}
	\end{figure}
	
	The scatter and binned analyses in \cref{Fig:5} show that OTOC similarity is threshold-like rather than a linear function of density overlap. While $\mathcal{S}(n,m)$ compares the static intensity profiles of PDFs, $Q_{c}(n,m)$ compares the complete time-dependent response of OTOC. In \cref{Fig:5}(a), pairs close to the density-overlap cut-off $\mathcal{S}(n,m)>0.61$ display a broad range of $Q_{c}$, showing that marginal spatial similarity alone is not sufficient to determine the OTOC response. In contrast, the largest-$\mathcal{S}$ pairs are strongly biased toward high $Q_{c}$ and low $d_{c}$. The binned median in \cref{Fig:5}(b) confirms this trend: increasing density-overlap similarity systematically enhances the median OTOC similarity and suppresses the median OTOC distance. Therefore, strongly overlapping eigenstate morphologies scramble information in nearly the same manner. In comparison, the residual spread arises from spectral phases, phase-space structure, symmetry constraints, and local matrix-element interference not captured by $\mathcal{S}(n,m)$ alone.
	
	\paragraph*{\label{sec:7}Conclusion---}
	In summary, we have shown that recurring eigenstate morphology provides a direct bridge between static nonthermal structure and dynamical scrambling in a chaotic quantum billiard. While entropy, participation ratios, and related scalar diagnostics identify anomalous localisation, they do not determine whether distinct eigenstates share the same spatial architecture. The normalised density overlap ($\mathcal{S}(n,m)$) resolves this missing structural information by grouping orthogonal eigenstates into morphology classes with a common intensity pattern. Comparing these classes with eigenstate-resolved OTOCs reveals that morphology has measurable dynamical consequences: strongly overlapping density profiles exhibit enhanced OTOC similarity ($Q_{c}(n,m)$) and suppressed curve distance ($d_{c}(n,m)$). The relation is threshold-like rather than purely linear, indicating that moderate spatial overlap is insufficient, whereas near-identical morphology strongly constrains short-to intermediate-time operator growth. 
	
	Our results show that scars are not merely static anomalies of eigenfunction intensity; when their morphologies recur, they also influence how quantum information scrambles. In conclusion, spatial morphology provides a structural constraint on scrambling.
	
	\paragraph*{Acknowledgments---}
	Numerical calculations were performed using \texttt{n2-highmem-32} virtual-machine instances on Google Cloud Platform, supported by Google Cloud research credits.
	
		\paragraph*{Data availability---}
	The data supporting this study are available from the corresponding author upon reasonable request and are intended to be deposited in a public repository upon publication.
	
	\bibliography{Bibliography}
\end{document}